%% file: 2022ISIThotplug-arxiv.tex
\documentclass[conference]{IEEEtran}

\usepackage[dvipsnames]{xcolor}
\usepackage{epsfig}
\usepackage{times}
\usepackage{float}
\usepackage{afterpage}
\usepackage{amsmath}
\usepackage{amstext,cite}
\usepackage{amssymb,bm}
\usepackage{latexsym}
\usepackage{color}
\usepackage{graphicx}
\usepackage{amsmath}
\usepackage{amsthm}
\usepackage{graphicx}
\usepackage[center]{caption}
\usepackage{pstricks}
\usepackage{caption}
\usepackage{subcaption} 
\usepackage{booktabs}
\usepackage{multicol}
\usepackage{lipsum}
\usepackage[T1]{fontenc}
\usepackage{hyperref}
\usepackage{aecompl}
\usepackage{mathrsfs}

\usepackage{soul}
\usepackage{cancel}

\allowdisplaybreaks

\input{macros}

\title{On Coded Caching Systems with Offline Users}

\begin{document}

\author{
\IEEEauthorblockN{Yinbin Ma and Daniela Tuninetti}
\IEEEauthorblockA{University of Illinois Chicago, Chicago, IL 60607, USA \\ Email:\{yma52, danielat\}@uic.edu}
}
\maketitle

\IEEEpeerreviewmaketitle

\begin{abstract}
Coded caching is a technique that leverages locally cached contents at the users to reduce the network's peak-time communication load.  
Coded caching achieves significant performance gains compared to uncoded caching schemes and is thus a promising technique to boost performance in future networks.
In the original model introduced by Maddah-Ali and Niesen (MAN), a server stores multiple files and is connected to multiple cache-aided users through an error-free shared link; once the local caches have been filled and {\it all} users have sent their demand to the server, the server can start sending coded multicast messages to satisfy all users' demands. 
A practical limitation of the original MAN model is that it halts if the server does not receive all users' demands, which is the limiting case of asynchronous coded caching when the requests of some users arrive with infinite delay.
In this paper we formally define a coded caching system where some users are offline. We propose achievable and converse bounds for this novel setting and show under which conditions they meet, thus providing an optimal solution, and when they are to within a constant multiplicative gap of two.
Interestingly, when optimality can be be shown, the optimal load-memory tradeoff only depends on the number active users, and not on the total (active plus offline) number of users.
\end{abstract}


\section{Introduction}
\label{sec:intro}

Coded caching, first introduced by Maddah-Ali and Niesen (MAN) in~\cite{maddah2014fundamental}, leverages locally cached contents at the users to reduce the communication load during peak-traffic times. A coded caching system has two phases. During the cache placement phase, the server populates the users' local caches, without knowing the users' future demands. During the delivery phase, the server broadcasts coded multicast messages to satisfy the users' demands. 
The achievable scheme proposed in~\cite{maddah2014fundamental} (referred to as MAN in the following) has combinatorial {\it uncoded cache placement} phase\footnote{Uncoded cache placement means that bits of the files are directly copied into the caches without coding.} and network coded delivery phase.
In~\cite{yu2017exact}, an improved delivery was proposed (referred to as YMA in the following), which improves on the MAN delivery by removing those linearly dependent multicast messages that occur when a file is requested by multiple users. The MAN placement with the YMA delivery meets with equality~\cite{yu2017exact} the converse bound derived in~\cite{wan2020index} under the constraint of uncoded placement; otherwise it is optimal within a factor of two~\cite{yu2018characterizing}. 

Coded placement strictly improves performance compared to uncoded placement, and can be exactly optimal. 
A non-exhaustive list of related works is as follows:
\cite{chen2016fundamental} showed how to achieve the cut-set bound in the small memory regime when there are more users than files; 
\cite{gomez2018fundamental} shows an improved performance compared to~\cite{chen2016fundamental} in the same regime;
\cite{tian2018symmetry} derived the optimal performance for the case of two users (and any number of files), and a partial characterization for the case of two files  (and any number of users).

A limitation of the classical coded caching setting~\cite{maddah2014fundamental} is that all users present during the placement phase must be active and synchronously send their demand 
before the delivery phase starts. 
The ``asynchronous demands'' setting, already discussed in~\cite{maddah2014fundamental}, allows the server to start transmission as soon as the first demand arrives; known schemes (as in~\cite{ghasemi2020asynchronous}, and references therein) however assume that all demands eventually arrive in finite time, otherwise the system fails to complete the delivery  
or the delivery time is infinite.

The case of coded caching with offline users is the focus of this paper. Here we assume that the demands of the offline users never arrive (or arrive with infinite delay) and the demands of the remaining users arrive synchronously. We refer to this setting as ``hotplug'' coded caching\footnote{Hotplug is a computer system term that refers to a device that can be added or removed from the running system without having to restart the system.}.
The ``decentralized'' coded caching setting, already discussed in~\cite{maddah2014fundamental}, allows each user to cache from the server at random and independently of the other users. This type of placement gives an achievable load for our hotplug setting because the decentralized scheme works for any number of user demands. In addition, the performance of the decentralized setting is useful to derive constant multiplicative gap results~\cite{maddah2014decentralized,yu2017exact}.

\paragraph*{Contributions}
In this paper, we first formalize the hotplug coded caching problem. Then, we propose two schemes that allow the demands of the active users to be satisfied regardless of the set of offline users, where the number (but not the identity) of the offline users is assumed known at the time of placement. Our schemes use coded cache placement.
\begin{enumerate}
\item
Our first new achievable scheme exploits Maximum Distance Separable (MDS) codes in the placement phase, where coding is done within each file but not across files. We show that such a strategy reduces the load significantly in the small cache size regime compared to a centralized baseline schemes. Furthermore, it achieves the optimal performance when the memory is small and the number of files is large. The matching converse is obtained from~\cite{yu2018characterizing}.
\item
Our second new achievable scheme applies MDS coding to the coded placement of~\cite{chen2016fundamental}. This scheme achieves the optimal performance when the memory is small and there are less files than users. The matching converse is obtained from the cut-set bound~\cite{maddah2014fundamental}.
\end{enumerate}

\paragraph*{Paper Outline}
The rest of the paper is organized as follows.
Section~\ref{sec: problem formation and known results} states the problem formulation and summarizes related known results.
Section~\ref{sec:main} summarizes our main results.
Section~\ref{sec:K'=N=2} shows the optimal scheme when there are two files and two active users. 
Section~\ref{sec:NumExamples} provides some numerical examples. 
Section~\ref{sec:conclusion} concludes the paper.
Some proofs can be found in Appendix.

\section{Problem Formulation and Known Results}
\label{sec: problem formation and known results}

\subsection{Notation Convention}
\label{sec: notation}
We adopt the following notation convention.
\begin{itemize}
    \item Calligraphic symbols denote sets, bold lowercase symbols %
    vectors, bold uppercase symbols matrices, and sans-serif symbols %
    system parameters.
    \item $\Mm[\Qc]$ denotes the submatrix of $\Mm$ obtained by selecting the rows indexed by $\Qc$.
    Similarly, $\dv[\Ic]$ is the subvector of $\dv$ obtained by selecting the elements indexed by $\Ic$.
    \item For an integer $b$, we let $[b] := \{1, \ldots, b\}$.
    \item For sets $\Sc$ and $\Qc$, we let $\Sc \setminus \Qc := \{k: k \in \Sc, k \notin \Qc\}$.
    \item For a vector $\dv$, $\mathsf{rank}(\dv)$ returns the number of distinct elements in $\dv$. For example, $\mathsf{rank}([1,5,5,1])=2$.
    \item For a ground set $\Gc$ and an integer $t$, we let $\Omega_{\Gc}^{t} := \{ \Tc \subseteq \Gc : |\Tc| = t\}$. 
    \item For $\Tc \in \Omega_{\Gc}^t$, we let $\Tc_i$ be the $i$-th subset in $\Omega_{\Gc}^t$ in lexicographical order. For example, the sets in $\Omega_{\{1,2,3\}}^2$ are indexed as $\Tc_1 = \{1,2\}$, $\Tc_2 = \{1,3\}$, $\Tc_3 = \{2,3\}$.
    \item For integers $a$ and $b$, $\binom{a}{b}$ is the binomial coefficient, or if  $a \geq b \geq 0$ does not hold.
  \end{itemize}

\subsection{Problem Formulation}
\label{sec: problem formation}

In a $(\Ksf, \Ksf^\prime, \Nsf)$ hotplug coded caching system:
\begin{itemize}
    \item A central server stores $\Nsf$ files, denoted as $F_1, \cdots, F_\Nsf$.
    \item Each file has $\Bsf$ i.i.d. uniformly distributed bits.
    \item The server communicates with $\Ksf$ users through a error-free shared link.
    \item Each user has a local memory that can contain up to $\Msf \Bsf$ bits, where $\Msf \in [0, \Nsf]$.
    We refer to $\Msf$ as the {\it memory size}. Caches are denoted as $Z_1, \ldots, Z_\Ksf$.
    \item The server sends the signal $X$ to the users through the shared link, where $X$ has no more than $\Rsf \Bsf$ bits, with $\Rsf \geq 0$.
    We refer to $\Rsf$ as the {\it load}. 
    \item The system has a placement phase and a delivery phase.
    The placement phase occurs at a time when the server is still unaware of which users will be active / not be offline, and which files the active users will request.
    We assume that the server knows that $\Ksf^\prime$ users will be active, with $\Ksf^\prime \leq \Ksf$. 
    The delivery phase occurs after the active users have sent their demands to the server.
\end{itemize}

In particular:

\paragraph*{Placement Phase} 
The server populates the local caches as a function of the files it stores, i.e.,
\begin{align}
    H(Z_k | F_1, \ldots F_\Nsf) = 0, \quad \forall k \in [\Ksf].
    \label{eq:cacheplacemnt}
\end{align}

\paragraph*{Delivery Phase} 
Once the set of active users becomes known to the server, denoted by $\Ic \in \Omega_{[\Ksf]}^{\Ksf^\prime}$, as well as the demand $d_k \in [\Nsf]$ of user $k\in\Ic$, the server starts sending. We denote the demands of all users by the vector $\dv = [d_1, \ldots d_\Ksf]$, thereby also including the demands of the offline users; the server is thus aware of the pair $(\Ic,\dv[\Ic])$. The message $X$ sent by the sever must guarantee that each active user, with the help of its locally cached content, can revere its desired file, i.e., for every $\Ic \in \Omega_{[\Ksf]}^{\Ksf^\prime}$ and $\dv[\Ic] \in [\Nsf]^{\Ksf^\prime}$, we must have
\begin{align}
    &H(X | \Ic, \dv[{\Ic}], F_1, \ldots F_\Nsf) = 0, 
    \label{eq:encoding}
  \\&H(F_{d_k} | Z_k, X) = 0, \quad \forall k \in \Ic.
    \label{eq:decoding}
\end{align}

\paragraph*{Performance} 
For $\Msf \in [0, \Nsf]$, we denote by $\Rsf^\star(\Msf)$ the minimum worst-case load, defined as
\begin{align}
&\Rsf^\star(\Msf) = \limsup_{\Bsf \rightarrow \infty} \ \min_{X,  Z_1, \ldots Z_\Ksf} \ \max_{\Ic,\dv[\Ic]} \{\Rsf: 
\notag\\&\quad
\textrm{\small conditions in~\eqref{eq:cacheplacemnt}-\eqref{eq:decoding} are satisfied with memory size $\Msf$}\}.
\end{align}

\subsection{Known Results for $\Ksf^\prime = \Ksf$}
When $\Ksf^\prime = \Ksf$, the hotplug model is equivalent to the classical setting in~\cite{maddah2014fundamental} for which the following is known. 

\paragraph*{MAN Placement Phase} 
Fix $t \in [0: \Ksf]$ and partition each file into $\binom{\Ksf}{t}$ equal-size subfiles as
\begin{align}
    F_i = ( F_{i,\Wc} \in \mathbb{F}_\qsf^{\Bsf / \binom{\Ksf}{t}} : \Wc \in \Omega_{[\Ksf]}^{t} ),  \quad \forall i \in [\Nsf].
    \label{eq:MANsplit}
\end{align}
For each user $k\in[\Ksf]$, the cache content $Z_k$ is
\begin{align}
    Z_k =( F_{i,\Wc} : i \in [\Nsf], \Wc \in \Omega_{[\Ksf]}^{t}, k \in \Wc ), \quad \forall k \in [\Ksf].
    \label{eq:MANacahe}
\end{align}
The memory size is $\Msf = \Nsf \binom{\Ksf-1}{t-1}/\binom{\Ksf}{t} = {\Nsf t}/{\Ksf}$.
The MAN placement is referred to a {\it centralized} as it requires coordination among users during the placement phase.

\paragraph*{MAN Multicast Messages} 
For the demand vector $\dv\in [\Nsf]^{\Ksf}$, the server constructs the multicast messages 
\begin{align}
    X_\Sc = \sum_{k \in \Sc} F_{d_k, \Sc \setminus \{k\}}, \quad \forall \Sc \in \Omega_{[\Ksf]}^{t+1}.
    \label{eq:MANmm}
\end{align}
Notice that user $k \in \Sc$ can recover the missing subfile $F_{d_k, \Sc \setminus \{k\}}$ from $X_\Sc$ in~\eqref{eq:MANmm} by ``caching out'' $\sum_{u \in \Sc \setminus \{k\}} F_{d_u, \Sc \setminus u}$ which can be computed from $Z_k$ in~\eqref{eq:MANacahe}.

\paragraph*{YMA Delivery Phase} 
In~\cite{yu2017exact} it was noted that some multicast messages in~\eqref{eq:MANmm} are linearly dependent on the others when a file is requested by multiple users.  
By not sending the redundant multicast messages, %
the lower convex envelope of the following points for all $t \in [0: \Ksf]$ is achievable 
\begin{align}
    ( \Msf_t, \Rsf^\text{\rm cen}_t )  = \left.\biggl(\Nsf \frac{\binom{\Ksf-1}{t-1}}{\binom{\Ksf}{t}}, \frac{\binom{\Ksf}{t+1} - \binom{\Ksf - r}{t+1}}{\binom{\Ksf}{t}} \biggr)\right|_{r=\min\{\Nsf,\Ksf\}}.
    \label{eq:performanceYMA}
\end{align}

\begin{rem}[Centralized vs. Decentralized] 
\rm \label{rem: centVsDecent}
Decentralized placement~\cite{maddah2014decentralized,yu2017exact} refers to the case where users cache each bit of the library i.i.d. at random with probability $\mu:=\Msf/\Nsf \in[0,1]$. 
An achievable memory-load tradeoff with such a decentralized placement is $\Rsf^\text{\rm de-cen}$ given by
\begin{align}
    \Rsf^\text{\rm de-cen}%
    = \left.\frac{1-\mu}{\mu}\Big( 1-(1-\mu)^r \Big)\right|_{\substack{\mu:={\Msf}/{\Nsf} \\ r:=\min\{\Nsf,\Ksf\}}} 
    \geq \Rsf^\text{\rm cen}, %
    \label{eq:performanceYMAdecentralized}
\end{align}
where $\Rsf^\text{\rm cen}$ is the lower convex envelop of~\eqref{eq:performanceYMA} and the inequality in~\eqref{eq:performanceYMAdecentralized} is from~\cite[eq(20)]{yu2018characterizing}.
For fixed $\mu$, $\Rsf^\text{\rm cen}$ depends on both $\Ksf$ and $r=\min\{\Nsf,\Ksf\}$, while $\Rsf^\text{\rm de-cen}$ only on $r$.
\hfill$\square$ \end{rem}

\section{Main Results}
\label{sec:main}
In this section we summarize our main results, which will be proved in the following sections.

\subsection{Achievability}
\label{sec:main results}
\begin{thm}[Achievability]
    \label{thm: achievable region}
    Let
    \begin{align} 
    \rsf^\prime := \min\{\Nsf,\Ksf^\prime\}.
    \end{align}
    For a $(\Ksf, \Ksf^\prime, \Nsf)$ hotplug system, %
    the lower convex envelope %
    of the following point is achievable
    \begin{align}
        ( \Msf_t, \Rsf^\text{\rm base}_t )  &= \biggl(\Nsf \frac{\binom{\Ksf-1}{t-1}}{\binom{\Ksf}{t}}, \frac{\binom{\Ksf}{t+1} - \binom{\Ksf - \rsf^\prime}{t+1}}{\binom{\Ksf}{t}} \biggr),         
    \label{eq:performanceBASE} %
    \forall t \in [0: \Ksf],
\\
        ( \Msf^\text{\rm new1}_t, \Rsf^\text{\rm new1}_t )  &=\biggl(\Nsf \frac{\binom{\Ksf-1}{t-1}}{\binom{\Ksf^\prime}{t}}, \frac{\binom{\Ksf^\prime}{t+1} - \binom{\Ksf^\prime - \rsf^\prime}{t+1}}{\binom{\Ksf^\prime}{t}} \biggr),
    \label{eq:performanceNEW1} %
    \forall t \in [0: \Ksf^\prime].
    \end{align}
When $\Ksf \geq \Ksf^\prime \geq \Nsf$, the following is achievable   
    \begin{align}
    ( \Msf^\text{\rm new2}, \Rsf^\text{\rm new2} ) = \biggl( \frac{1}{\Ksf^\prime}, \Nsf\left(1-\frac{1}{\Ksf^\prime}\right) \biggr).
    \label{eq:performanceNEW2}
\end{align}
\end{thm}

Few comments are in order.

\paragraph*{Baseline Scheme}
    The performance of the baseline scheme in~\eqref{eq:performanceBASE} %
    is that of a classical coded caching system with $\Ksf$ users and $\Nsf$ files but with a restricted set of demand vectors $\dv: \mathsf{rank}(\dv) \in [\min\{\Nsf,\Ksf^\prime\}]$, that is, the largest number of distinct files that can be requested is $\rsf^\prime =\min\{\Nsf,\Ksf^\prime\}$ (i.e., the minimum between the number of files and the number of {\it active} users) rather than $\min\{\Nsf,\Ksf\}$ (as in~\eqref{eq:performanceYMA}). Here the server ``fills in'' the demand of the offline users by repeating in a predefined order the demands of the active users\footnote{For example, the server does the YMA delivery as if the demand of each offline users is the same as the demand of the online user with the smallest index.} and uses the YMA delivery for the ``filled in'' demand vector; with this,  the number of distinct files that must be delivered by the server is not increased compared to that in the hotplug system.

\paragraph*{New Schemes}    
    Our first novel scheme attains the load in~\eqref{eq:performanceNEW1}--the proof can be found in Appendix~\ref{sec:NEW1achievablescheme}.  
    At a high level, we split each file into $\binom{\Ksf^\prime}{t}$ equal-length subfiles and then code the subfiles with an MDS code of rate ${\binom{\Ksf^\prime}{t}}/{\binom{\Ksf}{t}}$. 
    The placement of the MDS-coded symbols follows the MAN spirit and the delivery the YMA spirit.
    
    Our second novel scheme attains the load in~\eqref{eq:performanceNEW2}--the proof can be found in Appendix~\ref{sec:NEW2achievablescheme}. In this scheme, we first code the files together, and then we apply another level of MDS coding before the placement. The general delivery has two steps: the first one is to `decode' the cache contents of the active users as in~\cite{chen2016fundamental}, and the second one is to perform a sequence of YMA-same-file-deliveries to subsets of active users. 
    
\paragraph*{Comparisons}    
    In general $\Rsf^\text{\rm base} \leq \Rsf^\text{\rm new1} \leq \Rsf^\text{\rm de-cen}$ (all evaluated for $r=\rsf^\prime$), with $\Rsf^\text{\rm base} = \Rsf^\text{\rm new1} = \Rsf^\text{\rm cen}$ if $\Ksf = \Ksf^\prime$.

    By comparing the YMA load for a classical coded caching system with $\Ksf^\prime$ users and $r=\rsf^\prime =\min\{\Nsf,\Ksf^\prime\}$ in~\eqref{eq:performanceYMA}, with the load of our first new proposed scheme in~\eqref{eq:performanceNEW1}, we notice they have the exact same expression; the difference is in the memory requirement, which is 
    $\Msf/\Nsf %
    = t/\Ksf^\prime$ for the YMA scheme with $\Ksf^\prime$ users and 
    $\Msf/\Nsf %
    = t/\Ksf \cdot {\binom{\Ksf}{t}}/{\binom{\Ksf^\prime}{t}}$ for our first scheme with $\Ksf^\prime$ active users out of $\Ksf$ total users. In other words, we need more cache space (quantified by the inverse of the MDS code rate) in order to serve $\Ksf^\prime$ online users and tolerate $\Ksf - \Ksf^\prime$ offline users, compared to the classical YMA coded caching scheme for $\Ksf^\prime$ users. Note that the two schemes have the same memory requirement for $t=1$.

    Consider the following corner points 
\begin{align}
( \Msf^\text{\rm new1}_0, \Rsf^\text{\rm new1}_0 ) &= ( \Msf_0, \Rsf^\text{\rm base}_0 ) = (0, \rsf^\prime)|_{\rsf^\prime := \min\{\Nsf,\Ksf^\prime\}},
\label{eq:performanceNEWt=0}
 \\
( \Msf^\text{\rm new1}_1, \Rsf^\text{\rm new1}_1 ) &= 
\begin{cases}
\left(\frac{\Nsf}{\Ksf^\prime},\rsf^\prime-\frac{\rsf^\prime(\rsf^\prime+1)}{2\Ksf^\prime}\right) %
&  \Ksf^\prime-\rsf^\prime\geq 2
\\
\left(\frac{\Nsf}{\Ksf^\prime},\frac{\Ksf^\prime-1}{2}\right)
& \Ksf^\prime-\rsf^\prime\in\{ 0, 1\}
\end{cases}.
\label{eq:performanceNEW1:t=1}
\end{align}
The segment connecting the points~\eqref{eq:performanceNEWt=0} and~\eqref{eq:performanceNEW1:t=1} (achievable by memory sharing) outperforms the baseline scheme in the small memory regime and is optimal when the number of files is large enough; that connecting the points~\eqref{eq:performanceNEWt=0} and~\eqref{eq:performanceNEW2} is optimal in the small memory regime when the number of files is less than the number of users as stated in the next theorem.

\subsection{Optimality Guarantees} 
As a converse bound, we can use {\it any} converse result for the classical coded caching system with $\Ksf^\prime$ users and $\Nsf$ files; this is so because the performance of our hotplug system cannot be better than that of a system in which the server knows a priori which set of $\Ksf^\prime$ users will be active, and does the optimal placement and delivery for those users. With this type of converse bounds, we can show the following optimality result, whose proof can be found in Appendix~\ref{sec:converseproof}. 

\begin{thm}[Optimality Guarantees] 
\label{thm: optimality}
For a $(\Ksf, \Ksf^\prime, \Nsf)$ hotplug system.
We have the following optimality guarantees.
\begin{enumerate}

\item\label{item:opt:r=1}
When $\rsf^\prime=\min\{\Nsf, \Ksf^\prime\}=1$, $\Rsf^\text{\rm base}$ is optimal. 

\item\label{item:opt:K'=2N=2}
When $\Ksf \geq \Ksf^\prime = 2$ and $\Nsf = 2$, the optimal scheme has two non-trivial corner points: $(\Msf_1^\text{\rm new1},\Rsf_1^\text{\rm new1})=(1,1/2)$ and $(\Msf^\text{\rm new2},\Rsf^\text{\rm new2})=(1/2,1)$.

\item\label{item:opt:K'=2N>=3}
When $\Ksf \geq \Ksf^\prime = 2$ and $\Nsf \geq 3$, the only non-trivial optimal corner point is $(\Msf_1^\text{\rm new1},\Rsf_1^\text{\rm new1})=(\Nsf/2,1/2)$. 

\item\label{item:optNsmallMsmall-firstsegment}
When $\Nsf \leq \Ksf^\prime$ and $\Msf \leq \Nsf / \Ksf^\prime$, the corner point $(\Msf_1^\text{\rm new2},\Rsf_1^\text{\rm new2})=(1/\Ksf^\prime, \Nsf(1-1/\Ksf^\prime))$ is optimal.

\item\label{item:opt:NlargeMsmall-firstsegment}
When $\Nsf \geq \Ksf^\prime (\Ksf^\prime + 1)/2$ and $\Msf \leq \Nsf / \Ksf^\prime$,  the corner point $(\Msf_1^\text{\rm new1},\Rsf_1^\text{\rm new1})=(\Nsf/2,1/2)$ is optimal. 

\item\label{item:opt:lastsegment}
When $\Msf \geq \Nsf(1-1/\Ksf)$, $\Rsf^\text{\rm base}$ is optimal. 

\item\label{item:opt:gap is constant}
$\Rsf^\text{\rm base}$ is at most a factor 2 from optimal.

\end{enumerate}
\end{thm}

Few remarks are in order.

Item~\ref{item:opt:NlargeMsmall-firstsegment} with $\Ksf^\prime = 2$ only covers the first half of the memory range of Item~\ref{item:opt:K'=2N>=3}; the second half is not covered by Item~\ref{item:opt:lastsegment} as the memory regime in Item~\ref{item:opt:lastsegment} depends on $\Ksf$ (which can be any value no smaller than $\Ksf^\prime = 2$ in Item~\ref{item:opt:K'=2N>=3}). 

Theorem~\ref{thm: optimality} does not provide a tight characterization for $\rsf^\prime=\min\{\Nsf, \Ksf^\prime\}=2$ as the classical coded caching setting for two files is only partially solved~\cite{tian2018symmetry} (only up to three users).

The proof 
for Item~\ref{item:opt:r=1} and Item~\ref{item:opt:lastsegment} is as for the classical coded caching system, with converse given by the cut-set bound~\cite[Theorem 2]{maddah2014fundamental};
for Item~\ref{item:opt:K'=2N=2} is given in Section~\ref{sec:K'=N=2} and the converse is from~\cite{maddah2014fundamental};
for Item~\ref{item:opt:K'=2N>=3} the converse is~\cite[Theorem 3]{tian2018symmetry};
for Item~\ref{item:optNsmallMsmall-firstsegment} the converse is the cut-set bound~\cite[Theorem 2]{maddah2014fundamental}  %
for Item~\ref{item:opt:NlargeMsmall-firstsegment} the converse is~\cite[Theorem 2]{yu2018characterizing};
for Item~\ref{item:opt:gap is constant} uses~\cite[Lemma 1]{yu2018characterizing} (where one upper bounds the performance of the proposed centralized scheme by that of the decentralized one--see also Remark~\ref{rem: centVsDecent}; this is possible because the load of the classical coded caching model is bounded/finite when the number of users grows to infinity).

It is interesting to note that the exact optimality results in Theorem~\ref{thm: optimality} (except Item~\ref{item:opt:lastsegment}) do not depend on $\Ksf$ (the total number of users) but only on  $\Ksf^\prime$ (the total number of {\it active} users). It is not obvious that this should be the case in general.

\begin{figure}
    \centering 

   \begin{subfigure}[t]{0.4\textwidth}
   \centering 
   \includegraphics[width=0.9\textwidth]{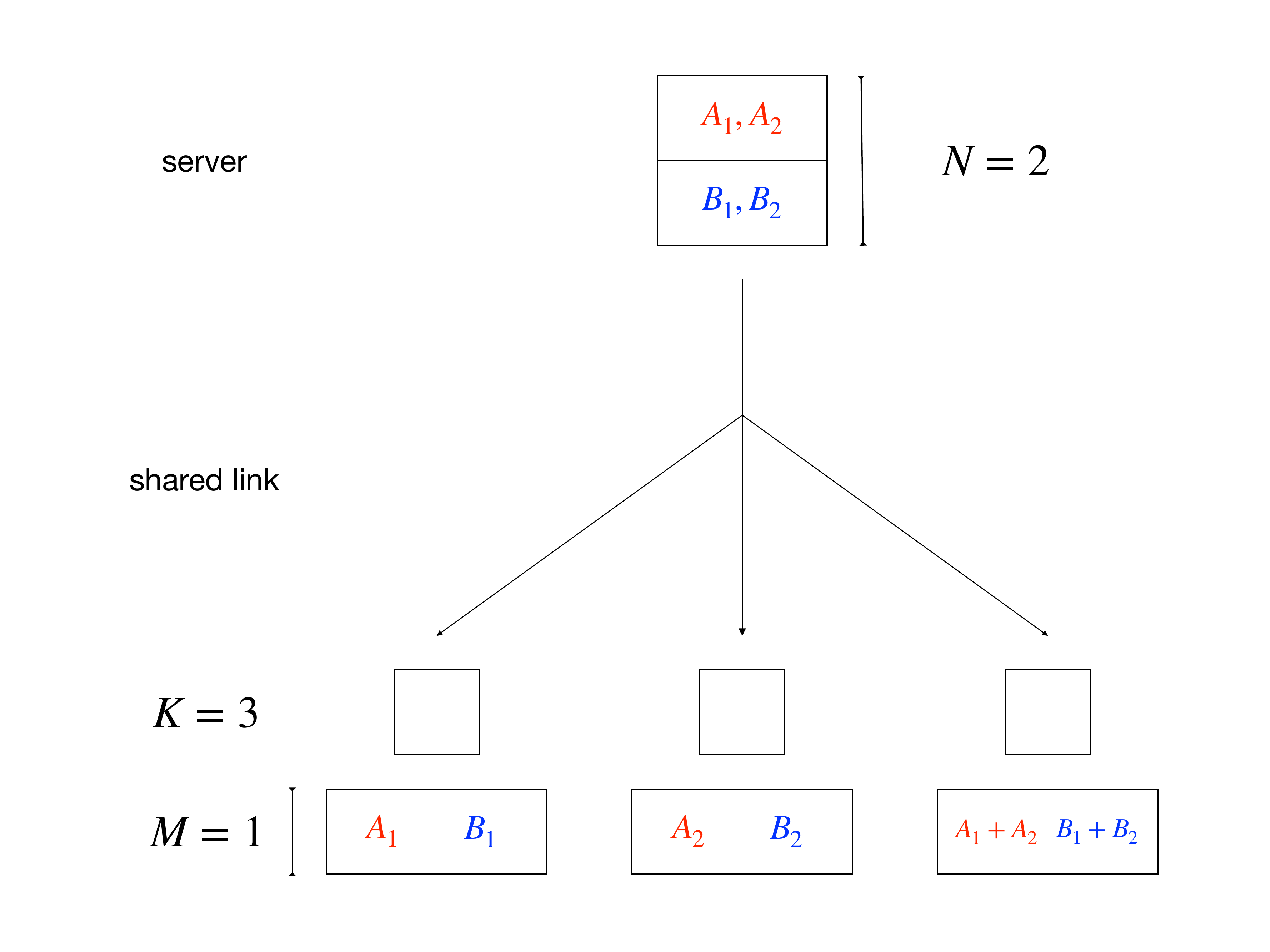}
    \caption{Cache contents for our first new scheme for memory size $\Msf=1$. The third user caches the two parity bits.} 
    \label{fig: caches (3,2,2)}
    \end{subfigure}

\vspace*{2mm}
\begin{subfigure}[t]{0.4\textwidth}
\centering 
\begin{align*}
\begin{array}{|l|c c c|}
\hline
            & \text{User~1} & \text{User~2} & \text{User~3} \\
            & \text{offline}& \text{offline}& \text{offline}\\
\hline
\dv=(1,1,1) & A_1 & A_2 &  A_1 + A_2 \\
\dv=(1,2,1) & A_1 + B_1 + B_2 & A_2 &  A_2 + B_1 \\
\hline
\end{array}
\end{align*}
\caption{The delivery for our first new scheme for memory size $\Msf=1$, for two different demand vectors as a function of which user is offline.}
\label{fig: signals (3,2,2)}
\end{subfigure}

    \begin{subfigure}[t]{0.4\textwidth}
    \centering 
    \includegraphics[width=\textwidth]{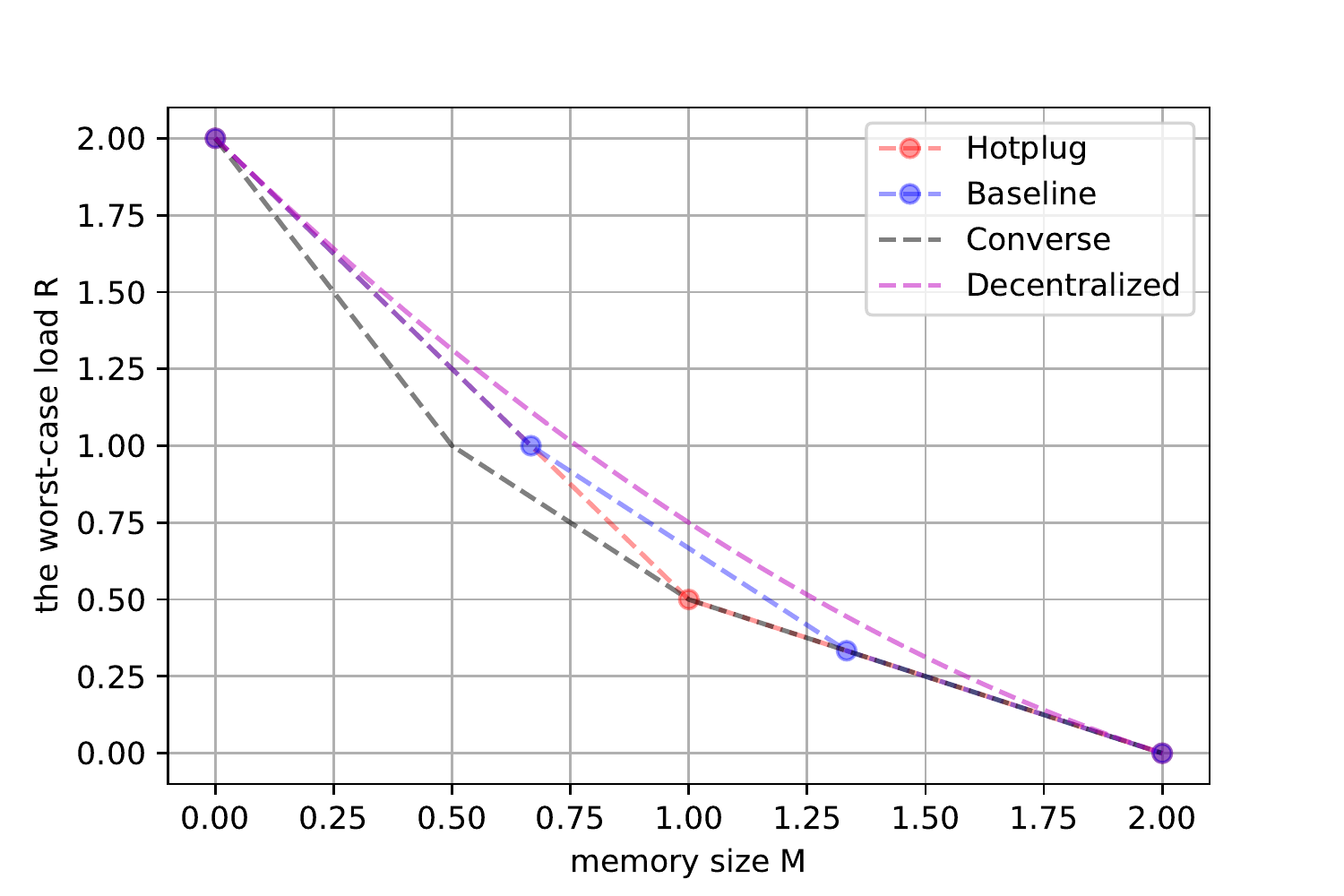} %
    \caption{Memory-load tradeoffs.}
    \label{fig: tradeoffs (3,2,2)}
    \end{subfigure}

    \caption{Memory-load tradeoffs for the hotplug system with $\Ksf=3$ users, $\Nsf=2$ files, and $\Ksf^\prime=2$ active users. 
    The converse is achievable for any $\Ksf \geq 3$. The performance of our new schemes does not depend on $\Ksf$.}
    \label{fig: hotplug (3,2,2)}

\end{figure}

\section{Optimality for $\Ksf \geq \Ksf^\prime = \Nsf = 2$}
\label{sec:K'=N=2}
We consider the hotplug system with $\Ksf\geq 3$ users, $\Nsf=2$ files, and $\Ksf^\prime=2$ active users, i.e., $\Ksf- \Ksf^\prime=1$ offline user. 

In this section we go into the proof details for $\Ksf=3$ users only, which is the simplest case that highlights the novelty of our new schemes.  The general case $\Ksf\geq 3$ follows from the proofs in Appendix~\ref{sec:NEW1achievablescheme} and~Appendix~\ref{sec:NEW2achievablescheme}.

Next, we aim to show the achievability of the two non-trivial corner points of the optimal region for the classical coded caching setting with two users and two files~\cite{maddah2014fundamental}, which is a converse bound for any hotplug system with $\Ksf \geq \Ksf^\prime = \Nsf = 2$.
To prove the achievability of the non-trivial corner points $(1,1/2)$ and $(1/2,1)$ (in addition to the trivial points $(2,0)$ and $(0,2)$) we proceed as follows.
We first derive the performance of our first new scheme, which achieves the point $(1,1/2)$ by using MDS coded placement (where coding is only within each file).
We then combine the coded placement idea of~\cite{maddah2014fundamental} with our MDS coded placement of our first new proposed scheme to show the achievability of the point $(1/2,1)$.

\paragraph*{Case $\Ksf=3$ and $\Msf=1$: First new scheme}
In Fig.~\ref{fig: hotplug (3,2,2)} we consider memory size $\Msf=1$ and $\Ksf=3$ users. 
The files are partitioned into two equal-size subfiles as $F_1=(A_1, A_2)$ and $F_2=(B_1, B_2)$. 
The subfiles of each file are coded with an MDS code of rate $2/3$. The cache contents are 
\begin{align*}
    Z_1 &= (A_1, B_1), \\
    Z_2 &= (A_2, B_2), \\
    Z_3 &= (A_1+A_2, B_1+B_2),
\end{align*}
as shown in Fig.~\ref{fig: caches (3,2,2)}. The third user caches the parity bits. 

Regardless of which user is active and what the other two demand, each active user must receive the missing half of the demanded file. Fig.~\ref{fig: signals (3,2,2)} gives the signals sent by the server according to Theorem~\ref{thm: achievable region}, for two different demand vectors as a function of which user is offline; all the other demand vectors can be dealt similarly. The load is $\Rsf_1^\text{\rm new1}=1/2$. 

Fig.~\ref{fig: tradeoffs (3,2,2)} shows the memory-load tradeoff attained by our first new scheme by the red dashed line, which is the lower convex envelope of the corner point $(1,1/2)$ achieved by the novel scheme with the trivial corner points $(0, 2)$ and $(2, 0)$. The blue dashed line represents the memory-load tradeoff when all three users are active. The gray dashed line is the optimal memory-load for a classical coded caching system with two users and two files~\cite{maddah2014fundamental}, which is achievable for any $\Ksf\geq 3$. For comparison, we also added to the figure the performance of a decentralized coded caching scheme in magenta dashed line, given by~\eqref{eq:performanceYMAdecentralized} with $r=\rsf^\prime=\min\{\Nsf,\Ksf^\prime\}=2$; the decentralized performance does not depend on $\Ksf$ and is an upper bound for the centralized performance for any $\Ksf$. 

This example shows that load savings are possible when the system is aware that only two users out of three can be active.

\begin{figure}
    \centering 
    \includegraphics[width=0.4\textwidth]{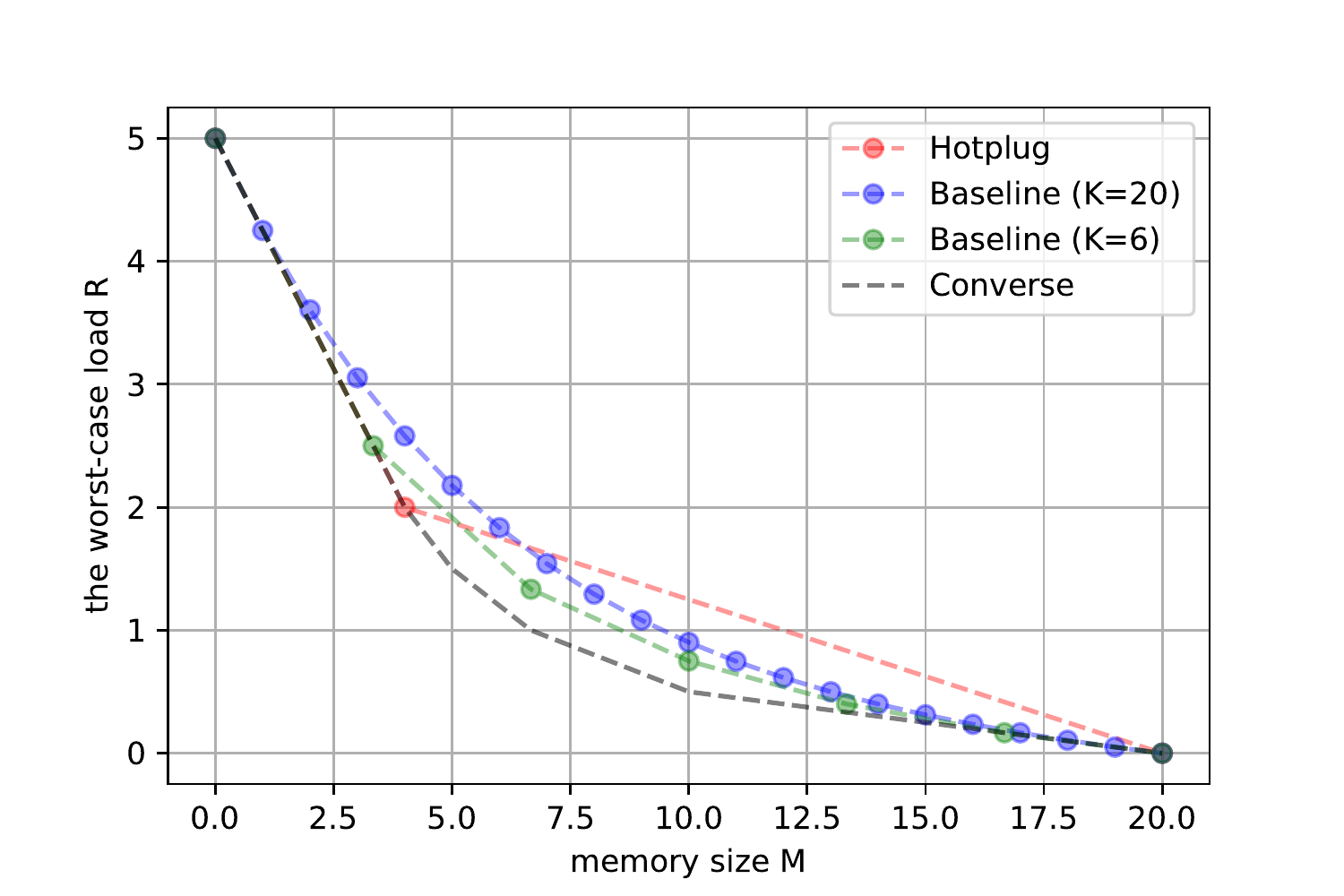}
    \caption{Memory-load tradeoffs for the hotplug system for with $(\Ksf^\prime, \Nsf) = (5, 20)$ and various values of $\Ksf$.}
    \label{fig: memory-tradeoff among various k}
\end{figure}

\paragraph*{Case $\Ksf=3$ and $\Msf=1/2$: Second new scheme}
Our first new scheme with MDS-coded placement attains only one corner point on the converse bound from~\cite{maddah2014fundamental}. 
In~\cite{maddah2014fundamental} it was shown that the point $(1/2,1)$ can be achieved by coded placement in the classical setting with two files and two users.
We next combine the idea of~\cite{maddah2014fundamental} with our MDS coded placement idea to show that $(1/2,1)$ is achievable for $\Ksf=3 > \Ksf^\prime=\Nsf=2$. 

Consider memory size $\Msf=1/2$ and $\Ksf=3$ users.  The files are partitioned as before %
but the cache contents are
\begin{align*}
    &\Am=[A_1; A_2], \ \Bm=[B_1; B_2], \ \text{(files seen as column vectors)}, \\
    &Z_1 = A_1+B_1 = \gv_1 (\Am+\Bm), \ \gv_1:=[1,0], \\
    &Z_2 = A_2+B_2 = \gv_2 (\Am+\Bm), \ \gv_2:=[0,1], \\
    &Z_3 = A_1+A_2 + B_1+B_2 = \gv_3 (\Am+\Bm), \ \gv_3:=[1,1].
\end{align*}
\label{eq:codedMANplacemnt}
When the pair of active users requests the same file, the server transmits the requested file.
 
For the pair of active users $(i,j)$ with $d_i=1,d_j=2$ the signal sent is 
\begin{align*}
X = ( \gv_j \Am, \ \gv_i \Bm ).
\end{align*}
User $i$ requesting file $\Am$  does
\begin{align*}
\begin{bmatrix} Z_i-\gv_i \Bm \\ \gv_j \Am \\ \end{bmatrix} 
= \underbrace{\begin{bmatrix} \gv_i   \\ \gv_j \\ \end{bmatrix}}_\text{$2 \times 2$ full rank matrix}  \Am,
\end{align*}
and similarly for user $j$ requesting file $\Bm$. 
Thus we can serve any pair of users, regardless of the demand, by $\Rsf^\text{\rm new2}=1$.

\paragraph*{Case $\Ksf\geq 3$} 
We showed that we can achieve all the corner points of the converse bound in~\cite{maddah2014fundamental} (which does not depends on $\Ksf$), thus we have the optimal coded caching strategy for the case $(\Ksf, \Ksf^\prime, \Nsf) = (3,2,2)$. 
The same approach extends to any $\Ksf\geq 3$ by using the general achievable schemes in Appendix~\ref{sec:NEW1achievablescheme} and Appendix~\ref{sec:NEW2achievablescheme}.

\begin{figure}
    \centering
    \begin{subfigure}[t]{0.4\textwidth}
        \centering 
        \includegraphics[width=\textwidth]{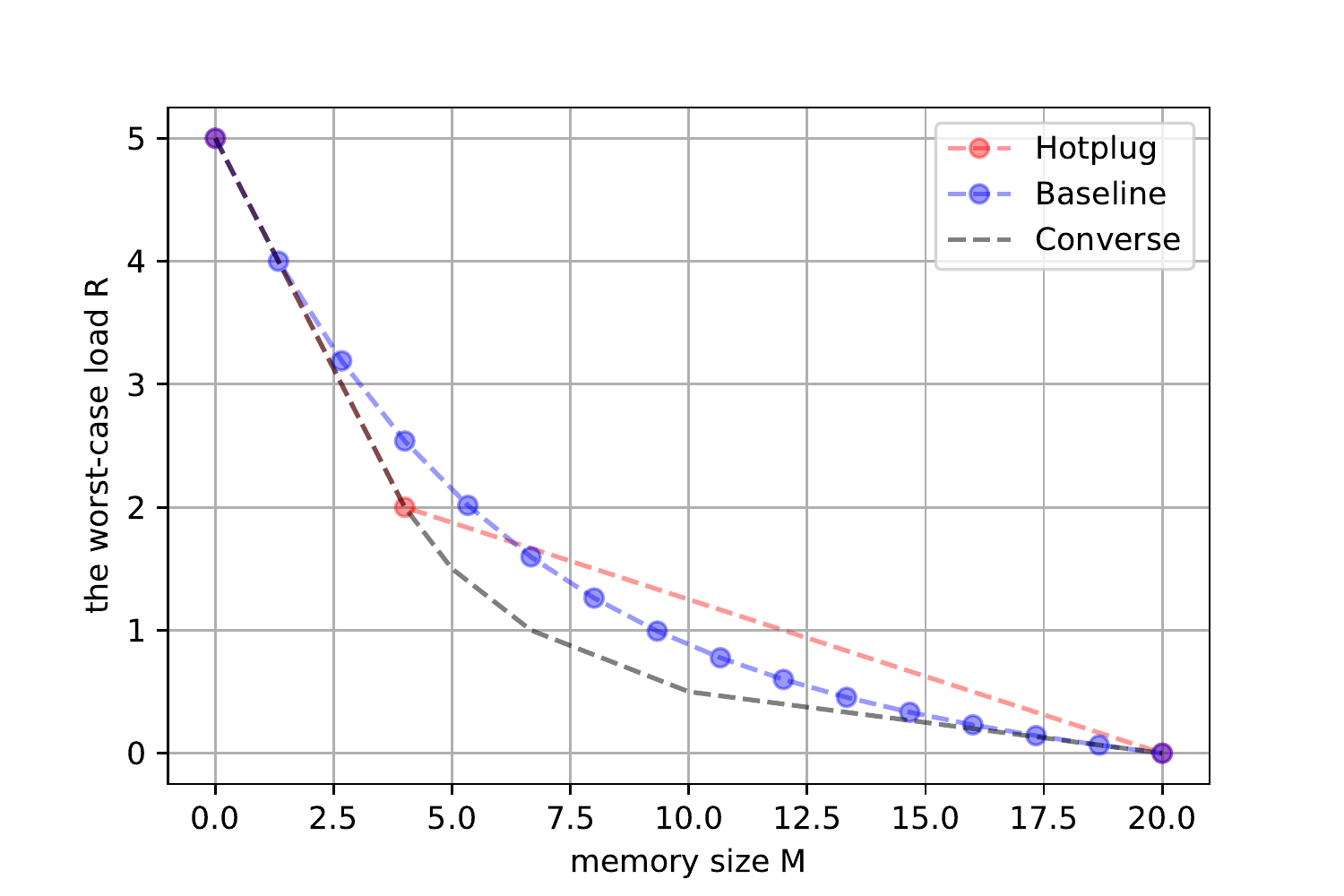} 
        \caption{Case $(\Ksf, \Ksf^\prime, \Nsf) = (10, 5, 20)$.} 
        \label{fig: memory-tradeoff for (10,5,20)}
    \end{subfigure}
    \begin{subfigure}[t]{0.4\textwidth}
        \centering 
        \includegraphics[width=\textwidth]{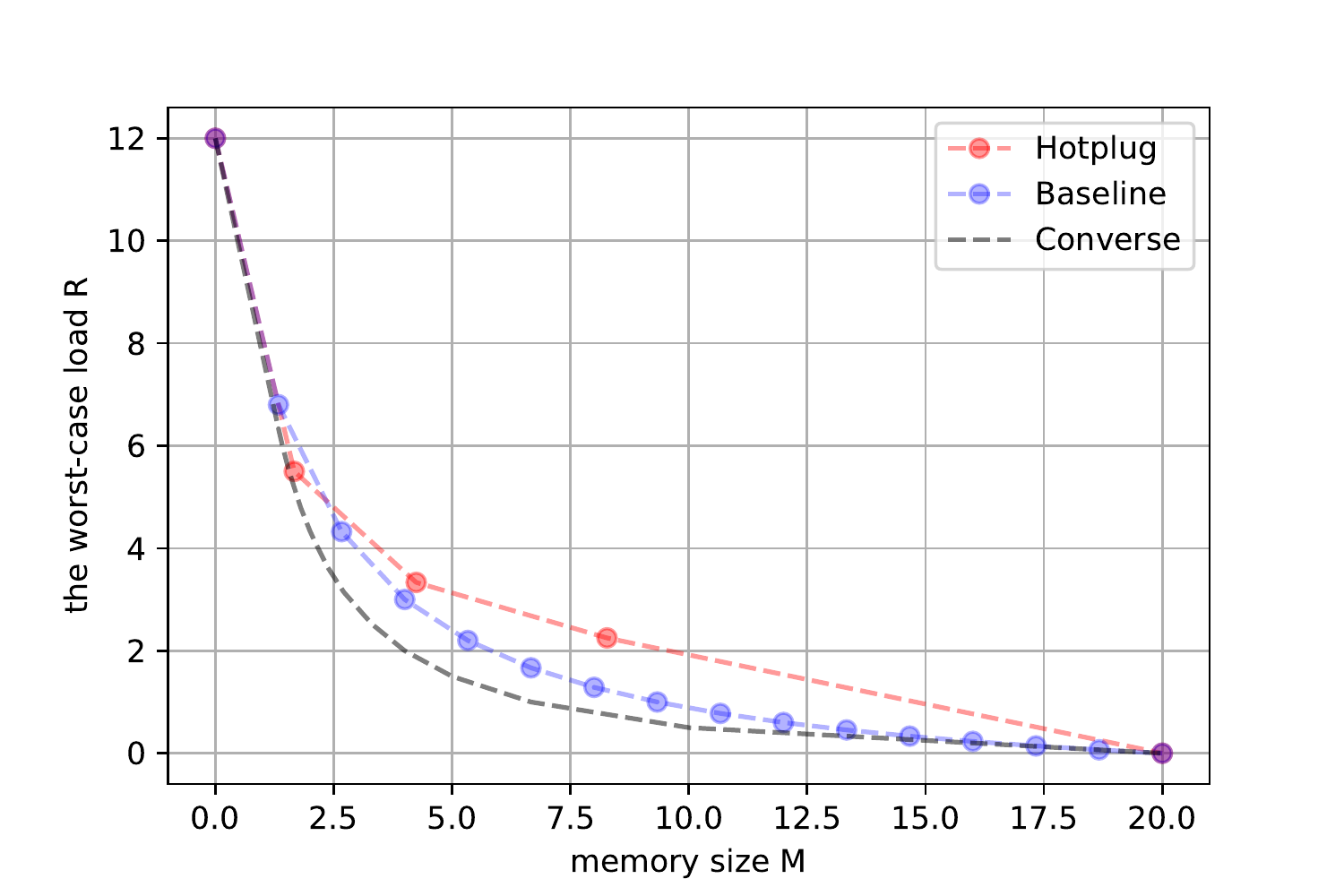}
        \caption{Case $(\Ksf, \Ksf^\prime, \Nsf) = (15, 12, 20)$.}
        \label{fig: memory-tradeoff for (15,12,20)}
    \end{subfigure}
    \caption{Memory-load tradeoffs for the hotplug system with $(\Ksf, \Nsf) = (15, 20)$ and different values of $\Ksf^\prime$.}
    \label{fig: memory-tradeoff for hotplug and baseline}
\end{figure}

\section{Numerical Evaluations}
\label{sec:NumExamples}
We conclude with some examples, to illustrate the performance of our new schemes.

{Case $(\Ksf^\prime, \Nsf) = (5, 20)$:}
Fig.~\ref{fig: memory-tradeoff among various k} shows the memory-load tradeoffs for the case $(\Ksf^\prime, \Nsf) = (5, 20)$ and various $\Ksf$.
The performance of the first new scheme and of the converse bound does not depend on the value of $\Ksf$, while that of the baseline scheme worsen as $\Ksf$ increases. 

{Case $(\Ksf, \Nsf) = (15, 20)$:}
Fig.~\ref{fig: memory-tradeoff for hotplug and baseline} shows the memory-load tradeoffs for two different values of $\Ksf^\prime$ for fixed $(\Ksf, \Nsf) = (15, 20)$. For $\Msf \in [0, \Nsf / \Ksf^\prime]$ in Fig.~\ref{fig: memory-tradeoff for (10,5,20)}, the first new scheme with MDS coded placement in Theorem~\ref{thm: achievable region} outperforms the baseline scheme in the small memory regime, and it is exactly optimal in the small memory regime.

\section{Conclusion}
\label{sec:conclusion}
In this paper, we introduced the novel hotplug coded caching model to address a practical limitation of the original coded caching system, namely,  to allow the server to start the delivery phase for a subset of active users, while the reaming users are offline. We proposed new coded caching schemes with MDS coded placement that are optimal in the small memory regime when some conditions hold. %
This shows that load savings are possible when the system is aware that only a subset of users will be active. 
Interestingly, when optimality can be be shown, the optimal performance only depends on the number active users.
Current work includes further extending optimality results.

This work was supported in part by NSF Award 1910309.

\clearpage 
\bibliographystyle{ieeetr}
\bibliography{references}

\clearpage 
\appendices

\section{Proof of Theorem~\ref{thm: optimality}}
\label{sec:converseproof}
In this section we are going to use the following results.

\begin{lem}[{Cut-set Bound from \cite[Theorem 2]{maddah2014fundamental}}] \label{lem: converse lemma MAN} \rm
For the classical coded caching system with with $\Ksf^\prime$ users and $\Nsf$ files, the memory-load pair $(\Msf, \Rsf)$ is lower bounded by
    \begin{align} \label{eq: load lower bound MAN}
        \Rsf\geq s - \frac{s}{\lfloor \Nsf/s \rfloor} \Msf, \quad \forall s \in [\min\{\Nsf, \Ksf^\prime\}].
    \end{align}
\end{lem}

\begin{lem}[{\cite[Appendix]{maddah2014fundamental}}] \label{lem: converse lemma MAN K'=N=2} \rm
For the classical coded caching system with with $\Ksf^\prime = 2$ users and $\Nsf=2$ files, the optimal memory-load is 
\begin{align} \label{eq: load lower bound MAN K'=N=2}
    \Rsf^\star = \max\left\{2-2\Msf, \frac{3}{2}-\Msf, 1-\frac{\Msf}{2}\right\}.
\end{align}
\end{lem}

\begin{lem}[{\cite[Theorem 3]{tian2018symmetry}}] \label{lem: converse lemma ChaoTian} \rm
For the classical coded caching system with with $\Ksf^\prime = 2$ users and $\Nsf \geq 3$ files, the optimal memory-load is %
\begin{align} \label{eq: load lower bound ChaoTian}
    \Rsf^\star = \max\left\{2-\frac{3\Msf}{\Nsf}, 1-\frac{\Msf}{\Nsf}\right\}.
\end{align}
\end{lem}

\begin{lem}[{\cite[Theorem 2]{yu2018characterizing}}] \label{lem: converse lemma YMA}\rm
For the classical coded caching system with with $\Ksf^\prime$ users and $\Nsf$ files, the memory-load pair $(\Msf, \Rsf)$ is lower bounded by
    \begin{align} \label{eq: load lower bound YMA}
        \Rsf \geq s - 1 + \alpha - \frac{s(s-1)-\ell(\ell-1)+2\alpha s}{2(\Nsf - \ell +1)} \Msf,
    \end{align}
    for any $s \in [\min\{\Nsf, \Ksf^\prime\}]$, $\alpha \in [0, 1]$, and where $\ell \in [s]$ is the minimum value such that
    \begin{align} \label{ieq: l condition}
        \frac{s(s-1)-\ell(\ell-1)}{2} + \alpha s \leq (\Nsf - \ell +1) \ell.
    \end{align}
\end{lem}

\subsection{Case~\ref{item:opt:r=1}: $\min\{\Nsf, \Ksf^\prime\}=1$}
In Theorem~\ref{thm: achievable region}, there are two (trivial) corner points $(0, 1)$ and $(\Nsf, 0)$. 
The line connecting these two points meets the lower bound in~\eqref{eq: load lower bound MAN} in Lemma~\ref{lem: converse lemma MAN} for $s=1$, namely $\Rsf \geq 1-\Msf/\Nsf$.
This optimal performance does not depend on the value of $\Ksf$.

\subsection{Case~\ref{item:opt:K'=2N=2}: $\Nsf = 2=\Ksf^\prime = 2$}
The achievability is proved in Section~\ref{sec:K'=N=2} and the converse is Lemma~\ref{lem: converse lemma MAN K'=N=2}.

\subsection{Case~\ref{item:opt:K'=2N>=3}: $\Nsf \geq 3,\Ksf^\prime=2$}
In Theorem~\ref{thm: achievable region}, there are three corner points points $(0, 2)$, $(\Nsf, 0)$ and $(\Nsf/2, 1/2)$. 
Their lower convex envelop equals the lower bound in~\eqref{eq: load lower bound ChaoTian} in Lemma~\ref{lem: converse lemma ChaoTian}.
This optimal performance does not depend on the value of $\Ksf$.

\subsection{Case~\ref{item:optNsmallMsmall-firstsegment}: small memory and fewer files than users}
The corner point 
$
(\Msf^\text{\rm new2},\Rsf^\text{\rm new2}) = (1/\Ksf^\prime, \ \Nsf(1-1/\Ksf^\prime)),
$
satisfies with equality the cut-set bound in Lemma~\ref{lem: converse lemma MAN} with $s=\Nsf$, namely  $\Rsf \geq \Nsf(1-\Msf)$, thereby showing that the segment connecting the trivial corner point $(\Msf,\Rsf) = (0,\Nsf)$ with this corner point is optimal.
This optimal performance does not depend on the value of $\Ksf$.

\subsection{Case~\ref{item:opt:NlargeMsmall-firstsegment}: small memory and many files}
When $\Nsf \geq \Ksf^\prime(\Ksf^\prime+1)/2$, the inequality in~\eqref{ieq: l condition} holds for any $s \in [\Ksf^\prime]$, $\alpha=1$ and $\ell=1$; so, if we let $s = \Ksf^\prime$ and $\Msf = \Nsf/\Ksf^\prime$ in~\eqref{eq: load lower bound YMA}, we obtain the lower bound
\begin{align}
    \Rsf \geq \Ksf^\prime - \frac{\Ksf^\prime(\Ksf^\prime+1)}{2} \ \frac{\Msf}{\Nsf}.
\label{eq:YMAconverse}
\end{align} 
The segment connecting the corner points $( \Msf^\text{\rm new1}_0, \Rsf^\text{\rm new1}_0 )=(0,\Ksf^\prime)$ and $( \Msf^\text{\rm new1}_1, \Rsf^\text{\rm new1}_1 ) =\biggl( \frac{\Nsf}{\Ksf^\prime}, \frac{\Ksf^\prime-1}{2} \biggr)$ satisfies~\eqref{eq:YMAconverse} with equality.
This optimal performance does not depend on the value of $\Ksf$.

\subsection{Case~\ref{item:opt:lastsegment}: large memory}
The segment connecting the corner points $( \Msf_{\Ksf-1}, \Rsf^\text{\rm base}_{\Ksf-1} )=(\Nsf\frac{\Ksf-1}{\Ksf}, \frac{1}{\Ksf})$ and $( \Msf_{\Ksf}, \Rsf^\text{\rm base}_{\Ksf} ) = (\Nsf, 0)$ satisfies $\Rsf \geq 1-\Msf/\Nsf$, which is Lemma~\ref{lem: converse lemma MAN} for $s=1$.

\subsection{Case~\ref{item:opt:gap is constant}: constant gap}
By Theorem~\ref{thm: achievable region} and Remark~\ref{rem: centVsDecent}, we have
\begin{align}
\Rsf^\star(\Msf)
      &%
       \leq \Rsf^\text{\rm base}(\Msf) 
       \leq \Rsf^\text{\rm de-cen}(\Msf) 
    \\&=\left.\frac{1-\mu}{\mu}\Big( 1-(1-\mu)^{\min\{\Nsf,\Ksf^\prime\}}\Big)\right|_{\mu:={\Msf}/{\Nsf}} 
    \\&\stackrel{\text{ \cite[Lemma 1]{yu2018characterizing} }}{\leq}  2.00884 \ \Rsf^\text{Lemma~\ref{lem: converse lemma YMA}},
\end{align}
where $\Rsf^\text{Lemma~\ref{lem: converse lemma YMA}}$ denotes the lower convex envelope of the region identified by Lemma~\ref{lem: converse lemma YMA}.

\section{Proof of Theorem~\ref{thm: achievable region}: First new scheme}
\label{sec:NEW1achievablescheme}

\paragraph*{Placement Phase} 
Fix $t \in [0: \Ksf^\prime]$ and partition each file into $\binom{\Ksf^\prime}{t}$ equal-size subfiles as 
\begin{align}
    F_i =  ( F_{i,\Wc} \in \mathbb{F}_\qsf^{\Bsf / \binom{\Ksf^\prime}{t}} : \Wc \in \Omega_{[\Ksf^\prime]}^{t} ),  \quad \forall i \in [\Nsf].
\end{align}
Then, for every $i \in [\Nsf]$, we treat the subfiles of each file as the information symbols of an MDS code with generator matrix $\Gm$ of dimension $\binom{\Ksf}{t} \times \binom{\Ksf^\prime}{t}$, i.e., any $\binom{\Ksf^\prime}{t}$ rows are linearly independent over $\mathbb{F}_\qsf$ for $\qsf$ a large enough prime number. 
The MDS-coded symbols are %
    \begin{align}
    \begin{bmatrix}
            C_{i,\Tc_1} \\ C_{i, \Tc_2} \\ \vdots \\ C_{i, \Tc_{\binom{\Ksf}{t}}}
        \end{bmatrix}
        &=\Gm \
        \begin{bmatrix}
            F_{i, \Wc_1} \\ F_{i, \Wc_2} \\ \vdots \\ F_{i, \Wc_{\binom{\Ksf^\prime}{t}}}
        \end{bmatrix}, \quad \forall i \in [\Nsf].
        \label{eq: mds subfiles}
    \end{align}

The cache contents are
\begin{align}
    Z_k = ( C_{i,\Tc}: i \in [\Nsf], \Tc \in \Omega_{[\Ksf]}^{t}, k \in \Tc ), \quad \forall k \in [\Ksf].
    \label{eq:NEWcache}
\end{align}
Thus the memory size is $\Msf^\text{\rm new1}_t$ as in Theorem~\ref{thm: achievable region}, which is the cache required by the MAN scheme for $\Ksf$ users divided by the rate of the MDS code used to `pre-code' each file.

\paragraph*{Delivery Phase} 
For any set of active users indexed by $\Ic \in \Omega_{[\Ksf]}^{\Ksf^\prime}$ with demands $\dv[\Ic] = [d_{i_1}, d_{i_2}, \ldots, d_{i_{\Ksf^\prime}}]$, the server forms the  the following multicast signals %
\begin{align}
    X_\Sc &= \sum_{k \in \Sc} C_{d_k, \Sc \setminus \{k\}}, \quad \forall \Sc\in \Omega_{\Ic}^{t+1}.
    \label{eq:NEWmm}
\end{align}
If the server were to broadcast all the multicast signals in~\eqref{eq:NEWmm}, %
the load would be $\binom{\Ksf^\prime}{t+1}/\binom{\Ksf^\prime}{t}$. 
Let $r = \mathsf{rank}\big(\dv[\Ic]\big)$; there are $\binom{\Ksf^\prime - r}{t+1}$ out of $\binom{\Ksf^\prime}{t+1}$ redundant multicast signals in~\eqref{eq:NEWmm}, which need not be sent (akin to the YMA delivery). For the largest possible $r = \rsf^\prime$, the load is $\Rsf^\text{\rm new1}_t$ as in Theorem~\ref{thm: achievable region}.

\paragraph*{Correctness} 
By leveraging the received multicast signal and the local cache content, that is, user $k\in\Ic$ knows $\big\{C_{d_k, \Qc}: \Qc \in \Omega_{\Ic}^{t} \big\}$, 
user $k\in\Ic$ has the following system of equations %
    \begin{align}
    \begin{bmatrix}
            C_{d_k,\Qc_1} \\ C_{d_k, \Qc_2} \\ \vdots \\ C_{d_k, \Qc_{\binom{\Ksf^\prime}{t}}}
        \end{bmatrix}
        &= \Gm[\Omega_{\Ic}^{\Ksf^\prime}] \
        \begin{bmatrix}
            F_{d_k, \Wc_1} \\ F_{d_k, \Wc_2} \\ \vdots \\ F_{d_k, \Wc_{\binom{\Ksf^\prime}{t}}}
        \end{bmatrix},
        \label{eq: mds recover}
    \end{align}
where $\Gm[\Omega_{\Ic}^{\Ksf^\prime}]$ is square and invertible by the properties the MDS code; thus, each active user obtains its desired file.

\begin{rem}
\rm \label{rem: just to prevent someone publishing it}

{\bf On extending the MAN+YMA scheme with MDS coding before placement.}
The following scheme achieves the same performance as our new first scheme for $\Msf=\Nsf/\Ksf^\prime$; after this memory value, it outperforms our new first scheme but is outperformed by the baseline scheme.

\paragraph*{Placement Phase} 
Let $\mathbf{G}_{\cdot}$'s be cache encoding matrices of size $\Bsf \eta \times \Bsf$, with $\eta$ to be determined later.
Let $t\in[\Ksf^\prime-1]$.
User $k\in[\Ksf]$ caches
\begin{align}
Z_k = ( \mathbf{G}_{\Tc} F_n : n\in[\Nsf], \Tc \in \Omega_{[\Ksf]}^{t}, k\in \Tc).
\end{align}
The cache size is
\begin{align}
\Msf = \Nsf \binom{\Ksf-1}{t-1}  \eta.
\end{align}

\paragraph*{Delivery Phase} 
For a given $(\Ic,\dv[\Ic])$, the server forms MAN-like multicast messages
\begin{align}
X_{\Sc} = \sum_{j\in\Sc} \mathbf{G}_{\Sc\setminus\{j\}} F_{d_j}, \ \Sc \in \Omega_{\Ic}^{t+1},
\end{align}
and sends them in YMA-fashion. 
The load is
\begin{align}
\Rsf = ( \binom{\Ksf^\prime}{t+1} - \binom{\Ksf^\prime - \rsf^\prime}{t+1})  \eta.
\end{align}

\paragraph*{Correctness} 
Each user gets from the server $\binom{\Ksf^\prime-1}{t}$ missing (coded)subfiles and has $\binom{\Ksf-1}{t-1}$ cached (coded)subfiles, thus decoding is possible if the collection of cache-coding matrices form an MDS matrix and
\begin{align}
( \binom{\Ksf^\prime-1}{t} + \binom{\Ksf-1}{t-1} ) \eta = 1.
\end{align}
Thus we achieve 
\begin{align}
(\Msf,\Rsf) = (\Nsf \frac{\binom{\Ksf-1}{t-1}}{\binom{\Ksf^\prime-1}{t} + \binom{\Ksf-1}{t-1}},  \frac{\binom{\Ksf^\prime}{t+1} - \binom{\Ksf^\prime - \rsf^\prime}{t+1}}{\binom{\Ksf^\prime-1}{t} + \binom{\Ksf-1}{t-1}} ),
\end{align}
which matches our first new coding scheme for $t=1$, but it is outperformed by the baseline scheme afterwards.

{\bf A specific example of a MAN+YMA scheme with MDS coding before placement which is optimal.}
The following is an example to show we can do better; part of current work is to generalize this idea.
Consider the hotplug subset with $\Ksf=6 > \Ksf^\prime=\rsf^\prime=3$.

\paragraph*{Placement Phase}
Partition each file into three parts seen as a column vector $F_n = [F_{n,1}; F_{n,2}; F_{n,3}]$. Consider binary cache-encoding matrices of size $2 \times 3$.
Cache 
\begin{align*}
Z_k &= (\mathbf{G}_{k} F_n :  n\in[\Nsf]), \quad \Msf/\Nsf = 2/3;
\\&\mathbf{G}_{1} = \begin{bmatrix} 1 & 0 & 0 \\ 0 & 1 & 0 \\ \end{bmatrix} = \begin{bmatrix}  \mathbf{g}_{12} \\ \mathbf{g}_{13} \\ \end{bmatrix};
\\&\mathbf{G}_{2} = \begin{bmatrix} 1 & 0 & 0 \\ 0 & 0 & 1 \\ \end{bmatrix} = \begin{bmatrix}  \mathbf{g}_{12} \\ \mathbf{g}_{23} \\ \end{bmatrix};
\\&\mathbf{G}_{3} = \begin{bmatrix} 0 & 1 & 0 \\ 0 & 0 & 1 \\ \end{bmatrix} = \begin{bmatrix}  \mathbf{g}_{13} \\ \mathbf{g}_{23} \\ \end{bmatrix};
\\&\mathbf{G}_{4} = \begin{bmatrix} \mathbf{g}_{13}\oplus \mathbf{g}_{23} \\ \mathbf{g}_{12}\oplus \mathbf{g}_{23} \\ \end{bmatrix};
\\&\mathbf{G}_{5} = \begin{bmatrix} \mathbf{g}_{13}\oplus \mathbf{g}_{23} \\ \mathbf{g}_{12}\oplus \mathbf{g}_{13} \\ \end{bmatrix};
\\&\mathbf{G}_{6} = \begin{bmatrix} \mathbf{g}_{12}\oplus \mathbf{g}_{23} \\ \mathbf{g}_{12}\oplus \mathbf{g}_{13} \\ \end{bmatrix};
\\& \text{define}
\\& \mathbf{g}_{14} = \mathbf{g}_{15} = \mathbf{g}_{16} = \mathbf{g}_{56} = \mathbf{g}_{12}\oplus \mathbf{g}_{13};
\\& \mathbf{g}_{25} = \mathbf{g}_{24} = \mathbf{g}_{26} = \mathbf{g}_{46} = \mathbf{g}_{12}\oplus \mathbf{g}_{23};
\\& \mathbf{g}_{36} = \mathbf{g}_{34} = \mathbf{g}_{35} = \mathbf{g}_{45} = \mathbf{g}_{13}\oplus \mathbf{g}_{23}.
\end{align*}

\paragraph*{Delivery Phase} 
For demand $(d_1,d_2,d_3)$ from users $u_1 < u_2 < u_3$, the server sends
\begin{align*}
X = \mathbf{g}_{u_2 u_3} F_{d_1}+\mathbf{g}_{u_1 u_3} F_{d_2}+\mathbf{g}_{u_1 u_2} F_{d_3}, \quad \Rsf = 1/3;
\end{align*}

\paragraph*{Correctness} 
Any three active users can decode as they have three linearly independent equations in three unknowns.

This scheme matches one corner point of the optimal memory-load tradeoff for the classical coded caching scheme with three users and three files~\cite{tian2018symmetry}.

\hfill$\square$ \end{rem}

\section{Proof of Theorem~\ref{thm: achievable region}: Second new scheme}
\label{sec:NEW2achievablescheme}

\begin{rem}
\rm \label{rem: just to easy the reader into the notation}

We start with the details of the case of two files to easy the reader into the notation.
We aim to show the achievability of the corner point
\begin{align}
(\Msf,\Rsf) = (1/\Ksf^\prime, \ 2(1-1/\Ksf^\prime)),
\end{align}
which satisfies with equality the cut-set bound in Lemma~\ref{lem: converse lemma MAN} with $s=2$, namely  $\Rsf \geq 2(1-\Msf)$, thereby showing that the segment connecting the trivial corner point $(\Msf,\Rsf) = (0,2)$ with this corner point is optimal.

\paragraph*{Placement Phase}The caches are populated as
\begin{align}
Z_k = \mathbf{G}_{k} (F_1+F_2), \ \forall k\in[\Ksf],
\end{align}
where $\mathbf{G}_{k}$ is the `cache-coding' matrix of user $k$ which is of dimension $\Bsf/\Ksf^\prime \times \Bsf$, thus $\Msf = 1/\Ksf^\prime$.

\paragraph*{Delivery Phase}
Consider a demand vector with $n_1$ active users demanding file $F_1$ and $n_2$ active users demanding file $F_2$, with $n_1+n_2=\Ksf^\prime \geq 2$.
When $n_1=0$ (or $n_2=0$), the server sends $F_1$ (or $F_2$) which has load $\Rsf=1 \leq 2(1-1/\Ksf)$.

Next we consider the case where both $n_1$ and $n_2$ are strictly positive.
Let $\Ic_1 \subseteq \Omega_{[\Ksf]}^{n_1}$ be the set of users demanding file $F_1$
and $\Ic_2 \subseteq \Omega_{[\Ksf]\setminus \Ic_1}^{n_2}$ be the set of users demanding file $F_2$.
The delivery has two steps.
\begin{enumerate}

\item
In the first step, the server's transmissions aim to `decode' the caches of the active users 
\begin{align}
X_\text{step1} = 
[\mathbf{G}_{j} F_1 : \forall j\in \Ic_2, \ 
 \mathbf{G}_{i} F_2 : \forall i\in \Ic_1
]. 
\end{align}
There are $n_1+n_2=\Ksf^\prime$ sub-messages in $X_\text{step1}$, each of size  $\Bsf/\Ksf^\prime$.
The net result of this first step is that the active users have now an `decoded' cache containing
\begin{align}
Z_i^\prime &= (\mathbf{G}_{u} F_1, \ \forall u\in \Ic_2 \cup \{i\}), \ \forall i\in \Ic_1,
\\
Z_j^\prime &= (\mathbf{G}_{u} F_2, \ \forall u\in \Ic_1 \cup \{j\}), \ \forall j\in \Ic_2.
\end{align}

\item
In the second step, the server creates MAN-type multicast messages to serve pairs of active users requesting the same file. For any two users in $\Ic_1$ (or in $\Ic_2$), we face a classical MAN problem with $(t,r)=(1,1)$ where each subfile is cached exclusively by one user and all the users request the same file. Thus, with $u^\star_n = \min\{u : u \in \Ic_n\}$ being the `leader' user for file $n\in[2]$, we have
\begin{align}
X_\text{step2} = 
[&\mathbf{G}_{u^\star_1} F_1 + \mathbf{G}_{i} F_1 : \forall i\in \Ic_1\setminus\{u^\star_1\}, \\
 &\mathbf{G}_{u^\star_2} F_2 + \mathbf{G}_{j} F_2 : \forall j\in \Ic_2\setminus\{u^\star_2\}
]. 
\end{align}
There are $n_1-1+n_2-1=\Ksf^\prime-2$ sub-messages in $X_\text{step2}$, each of size  $\Bsf/\Ksf^\prime$.
\end{enumerate}

In total the server has sent $\Ksf^\prime-2+\Ksf^\prime=2(\Ksf^\prime-1)$ sub-messages, each of size $\Bsf/\Ksf^\prime$. 
The load is thus $\Rsf=2(1-/\Ksf^\prime)$, as claimed.

\paragraph*{Correctness}
We still need to show that each active user can decode its demanded file.
At the end of the delivery phase, each active user $k \in \Ic_1 \cup \Ic_2$ (recall $ |\Ic_1 \cup \Ic_2| = \Ksf^\prime$) has the following set of equations
\begin{align}
\underbrace{[
 \mathbf{G}_{u} : \forall u \in \Ic_1 \cup \Ic_2
]}_\text{$\Ksf^\prime \frac{\Bsf}{\Ksf^\prime} \times \Bsf$ matrix},
\underbrace{F_{d_k}}_\text{$\Bsf \times 1$ vector}
\end{align}
which can be inverted if the collection of cache-encoding matrices $\{ \mathbf{G}_{1}, \mathbf{G}_{2}, \ldots \mathbf{G}_{\Ksf}\}$ has the following MDS-like property: every $[\mathbf{G}_{\Sc} : \forall \Sc \in \Omega_{[\Ksf]}^{\Ksf^\prime}]$ is full rank. Such matrices exists.
\hfill$\square$ \end{rem}

We aim to show the achievability of the corner point
\begin{align}
(\Msf,\Rsf) = (1/\Ksf^\prime, \ \Nsf(1-1/\Ksf^\prime)).
\end{align}

\paragraph*{Placement Phase}
The caches are populated as
\begin{align}
Z_k = \mathbf{G}_{k} (F_1+F_2+\ldots+F_{\Nsf}), \ \forall k\in[\Ksf],
\end{align}
where $\mathbf{G}_{k}$ is the `cache-coding' matrix of user $k$ which is of dimension $\Bsf/\Ksf^\prime \times \Bsf$, thus $\Msf = 1/\Ksf^\prime$.

\paragraph*{Delivery Phase}
Consider a demand vector with $n_j$ active users demanding file $F_j$ for $j\in[\Nsf]$, with $n_1+n_2+\ldots+n_{\Nsf}=\Ksf^\prime \geq \Nsf$.
When at least one of the $n_j$ is zero, the server sends all the demanded files, which has load $\Rsf \leq \Nsf-1 \leq \Nsf(1-1/\Ksf)$.

Next we consider the case where all $n_j$'s are strictly positive.
Let $\Ic_j \subseteq \Omega_{[\Ksf]}^{n_j}$ be the set of active users demanding file $F_j$, where the $\Ic_j$'s are disjoint and $|\Ic_j|=n_j$ for $j\in[\Nsf]$.
The delivery has two steps.
\begin{enumerate}

\item
In the first step, the server's transmissions aim to `decode' the caches of the active users 
\begin{align}
X_\text{step1} = 
[ 
&\mathbf{G}_{u} F_n : \forall u\in \Ic_1, \ n\in[\Nsf]\setminus\{1\}, \\
&\mathbf{G}_{u} F_n : \forall u\in \Ic_2, \ n\in[\Nsf]\setminus\{2\}, \\
&\vdots \\
&\mathbf{G}_{u} F_n : \forall u\in \Ic_{\Nsf}, \ n\in[\Nsf]\setminus\{\Nsf\}
]. 
\end{align}
There are $\Ksf^\prime(\Nsf-1)$ sub-messages in $X_\text{step1}$, each of size  $\Bsf/\Ksf^\prime$.

Let $\Ic = \cup_{j\in[\Nsf]} \Ic_j$, with $|\Ic|=n_1+n_2+\ldots+n_{\Nsf}=\Ksf^\prime$.
The net result of this first step is that the active users have now an `unlocked' cache containing
\begin{align}
Z_i^\prime &= (\mathbf{G}_{u} F_\ell, \ \forall u\in (\Ic \setminus \Ic_\ell) \cup\{i\}),  \forall i\in \Ic_\ell, \ \ell\in[\Nsf],
\end{align}

\item
In the second step, the server creates MAN-type multicast messages to serve pairs of active users requesting the same file. 
For any two users in $\Ic_n$, we face a classical MAN problem where each subfile is cached exclusively by one user in $\Ic_n$ and all the users in $\Ic_n$ request the same file $F_n$, for $n\in[\Nsf]$.
Thus, with $u^\star_n = \min\{u : u \in \Ic_n\}$ being the `leader' user for file $n\in[\Nsf]$, we have
\begin{align}
X_\text{step2} = 
[&\mathbf{G}_{u^\star_1} F_1 + \mathbf{G}_{j} F_1 : \forall j\in \Ic_1\setminus\{u^\star_1\}, \\
 &\mathbf{G}_{u^\star_2} F_2 + \mathbf{G}_{j} F_2 : \forall j\in \Ic_2\setminus\{u^\star_2\}, \\
 &\vdots  \\
 &\mathbf{G}_{u^\star_{\Nsf}} F_{\Nsf} + \mathbf{G}_{j} F_{\Nsf} : \forall j\in \Ic_{\Nsf}\setminus\{u^\star_{\Nsf}\}
]. 
\end{align}
There are $\Ksf^\prime-\Nsf$ sub-messages in $X_\text{step2}$, each of size  $\Bsf/\Ksf^\prime$.
\end{enumerate}

In total the server has sent $\Ksf^\prime(\Nsf-1) + \Ksf^\prime-\Nsf =\Nsf(\Ksf^\prime-1)$ sub-messages, each of size $\Bsf/\Ksf^\prime$. 
The load is thus $\Nsf(1-/\Ksf^\prime)$, as claimed.

\paragraph*{Correctness}
At the end of the delivery phase, each active user $k \in \Ic$ has the following set of equations
\begin{align}
\underbrace{[
 \mathbf{G}_{u} : \forall u \in \Ic
]}_\text{$\Ksf^\prime \frac{\Bsf}{\Ksf^\prime} \times \Bsf$ matrix},
\underbrace{F_{d_k}}_\text{$\Bsf \times 1$ vector}
\end{align}
which can be inverted if the collection of cache-encoding matrices $\{ \mathbf{G}_{1}, \mathbf{G}_{2}, \ldots \mathbf{G}_{\Ksf}\}$ has the following MDS-like property: every $[\mathbf{G}_{\Sc} : \forall \Sc \in \Omega_{[\Ksf]}^{\Ksf^\prime}]$ is full rank. Such matrices exists on a large enough finite field.

\end{document}

%% file: macros.tex
\long\def\comment#1{}

\newcommand{\dv}{{\mathbf d}}

\newcommand{\gv}{{\mathbf g}}

\newcommand{\Am}{{\mathbf A}}
\newcommand{\Bm}{{\mathbf B}}

\newcommand{\Gm}{{\mathbf G}}

\newcommand{\Mm}{{\mathbf M}}

\newcommand{\Gc}{{\mathcal G}}

\newcommand{\Ic}{{\mathcal I}}

\newcommand{\Qc}{{\mathcal Q}}

\newcommand{\Sc}{{\mathcal S}}
\newcommand{\Tc}{{\mathcal T}}

\newcommand{\Wc}{{\mathcal W}}

\newcommand{\qsf}{{\mathsf q}}
\newcommand{\rsf}{{\mathsf r}}

\newcommand{\Bsf}{{\mathsf B}}

\newcommand{\Ksf}{{\mathsf K}}

\newcommand{\Msf}{{\mathsf M}}
\newcommand{\Nsf}{{\mathsf N}}

\newcommand{\Rsf}{{\mathsf R}}

\theoremstyle{definition}
\newtheorem*{rem*}{Remark}

\theoremstyle{plain}
\newtheorem{thm}{Theorem}%

\newtheorem{lem}{Lemma}

\newtheorem{rem}{Remark}

\providecommand{\definitionname}{Definition}